\documentstyle[prl,aps,preprint]{revtex}
\begin{document}
\draft
\flushbottom
 
\title{Spin-Polarized Transport Across an La$_{0.7}$Sr$_{0.3}$MnO$_{3}$/YBa$_{2}$Cu$_{3}$O$_{7-x}$ Interface: Role of Andreev Bound States}

\author{Z. Y. Chen\cite{ee}, Amlan Biswas, Igor \v{Z}uti\'c, T. Wu, 
S. B. Ogale, R. L. Greene, and T. Venkatesan\cite{ee}}

\address{Center for Superconductivity Research, Department of Physics, University of Maryland, College Park, Maryland 20742}

\date{\today}
\maketitle

\begin{abstract}
Transport across an La$_{0.7}$Sr$_{0.3}$MnO$_{3}$/YBa$_{2}$Cu$_{3}$O$_{7-x}$
(LSMO/YBCO) interface is studied as a function of temperature and surface 
morphology. For comparison, control measurements are performed in non-magnetic 
heterostructures of  LaNiO$_{3}$/YBCO and Ag/YBCO.
In all cases, YBCO is used as bottom layer to eliminate the channel resistance
and to minimize thermal effects. The observed differential conductance reflects 
the role of Andreev bound states in $a-b$ plane, and brings out for the first
time the suppression of such states by the spin-polarized transport across the 
interface. The theoretical analysis of the measured data reveals decay of the 
spin polarization near the LSMO surface with temperature, consistent with the
reported photoemission 
data.
\end{abstract}
\newpage

The pioneering work on spin-polarized transport in conventional 
superconductors~\cite{Mes94}, performed in the tunneling limit of strong
interfacial scattering, has been recently extended to high transparency
ferromagnet/superconductor heterostructures~\cite{Jon95,Sou98}, where the 
two-particle process of Andreev reflection plays an important role. There is  
generally a good understanding of transport in normal 
metal/high temperature superconductor (HTSC) junctions and the 
important role of the Andreev bound states (midgap states)~\cite{Hu94,Cov97,Dag00,Fog97,Wal99},
which lead to the formation of the zero bias conductance peak (ZBCP).
In contrast, 
there remain many open questions in the spin-polarized case involving 
heterostructures of colossal magnetoresistance materials (CMR) and 
HTSC~\cite{Vas97,Vas98,Zhu99,Zut00}.
Systematic studies of these systems with high spin polarization
are particularly important as they hold potential for probing ferromagnetic interfaces,
unconventional superconductivity~\cite{Zhu99,Zut00} and for 
modifying superconducting properties, such as the critical current 
and T$_c$~\cite{Vas97}. Understanding and control of interface properties in CMR/HTSC 
heterostructures could also lead to novel spin-based devices.
The key factors in such studies are the electronic and magnetic quality of the interface 
region on the sides of the ferromagnet and the superconductor, and the thermal 
management of the device configuration, since these can critically influence the outcome
of the measurement.  

In this work we report and analyze some observations of spin-polarized
transport at a high quality 
La$_{0.7}$Sr$_{0.3}$MnO$_{3}$/YBa$_{2}$Cu$_{3}$O$_{7-x}$ (LSMO/YBCO) interface. 
In contrast to previous studies~\cite{Vas98,Saw99}, YBCO is used as the bottom
electrode. This choice of geometry (Fig.~\ref{fig1}a)
eliminates the channel resistance and
minimizes heating effects, thereby unfolding some peculiar features in the 
differential conductance-voltage (G-V) characteristics, not reported so far.
By analyzing these features we are able to show that the Andreev bound states, 
observed in the $a-b$ plane HTSC tunneling experiments~\cite{Cov97,Dag00},
have a major influence on the transport properties across the CMR/HTSC interface.
These results also reveal, for the first time, the suppression of such bound 
states by the spin-polarized transport across the interface, as predicted
theoretically~\cite{Zhu99,Zut00}. 
Our analysis of the interfacial spin polarization is consistent with the
findings of photoemission experiments, which show that the surface spin 
polarization of LSMO decreases more rapidly with temperature
than that of the bulk~\cite{Par98}. 

All films involved in the measurement were made by the pulsed laser deposition 
(PLD) technique. Before we address the issue of high quality film growth
 it is helpful to 
discuss our testing structure, shown in Fig.~\ref{fig1}a.
Several important considerations were involved in fabricating this seemingly 
simple structure to achieve high quality and reproducibility of the results. 
In particular, the use of YBa$_{2}$Cu$_{3}$O$_{7-x}$ as the bottom layer
offers three significant advantages: a) eliminating the channel resistance 
contribution at temperatures below the T$_{c}$ of YBCO. Thus, the CMR/HTSC
interface contribution dominates the resistance; b) providing an equipotential 
surface leading to a uniform and perpendicular current distribution;
and c) avoiding heating effects from the CMR layer, which is significant 
in other structures.
Au pads of 150 $\times$ 200 $\mu$m
separated by 500 $\mu$m were patterned using optical lithography, and ion 
milled down to the YBCO layer in Argon ambient~\cite{Jsi}. In order to ensure
absence of possible shorting, the YBCO layer was always 15$\%$ over-etched. 
25 $\mu$m-thick gold wires were directly bonded to the top gold contact
layer for measurements of current-voltage characteristic in a four-probe 
configuration. The results of such measurements were differentiated digitally 
to extract G-V curves. 
In the PLD procedure: first, a 2000 \AA~ film of c-axis YBCO was grown
on a (100) SrTiO$_{3}$ substrate using 
an energy density of 1.7 J/cm$^{2}$ in O$_{2}$ pressure of 150 mtorr at
800 $^{\circ}$C giving a T$_{c}$ of 90 K. A 700 \AA~ film of LSMO was then
deposited on the YBCO layer using an energy density of 2 J/cm$^{2}$ in 
400 mtorr O$_{2}$ at 780 $^{\circ}$C and the structure was allowed to
cool naturally in 400 torr O$_{2}$. Finally, 200 \AA~ gold was deposited 
{\it in  situ} in a vacuum of 1$\times 10^{-6}$ torr at 90 $^{\circ}$C to
protect the surface, followed by another {\it ex  situ} deposition of a 
3000 \AA~ Au layer by thermal evaporation. Rocking curves routinely have
full width of half maximum of less than 0.3$^{\circ}$ for both YBCO 
and LSMO peaks, indicating the high quality of the films produced by this
procedure. After all the processing steps the T$_{c}$ of the YBCO films 
remains 90 K. 
We performed a similar experiment on much smoother YBCO films, specifically 
grown under slightly off-optimum conditions~\cite{Off} (substrate temperature:
780 $^{\circ}$C, energy density: 1.4 J/cm$^{2}$). Although the T$_{c}$ in 
this case was only 7 K lower, the surface morphology was
considerably flatter, as shown in the insets of Fig.~\ref{fig1}b and 1c.

The resistance of the LSMO/YBCO junction is composed of three parts: the
Au/LSMO interface, the LSMO electrode and the LSMO/YBCO interface. 
According to the literature~\cite{Mie98}, even an {\it ex situ} grown
Au/LSMO interface has a surface resistivity of about 1$\times$10$^{-6}$ 
$\Omega$ cm$^{2}$ at room temperature, hence the resistance coming from 
the Au/LSMO interface
in our case is smaller than 3.3 m$\Omega$. Taking a very   
conservative 
estimate for 
the resistivity of LSMO of ~10 m$\Omega$ cm, we conclude that the LSMO
electrode has a resistance of about 2.3 m$\Omega$. Given that the total 
resistance in our structure is about 300 m$\Omega$, these parasitic
resistances can be neglected and the G-V curves below T$_{c}$ can be
regarded as genuinely 
representing the properties of the LSMO/YBCO interface. 

The conductance data, shown for the ``faceted'' (Fig.~\ref{fig1}b)~\cite{field}
and smooth
(Fig.~\ref{fig1}c) interface, display a general ``V''-shaped background,
similar to the tunneling data in metallic oxide systems~\cite{Sun98,Ray95}. 
The G-V curves were seen to converge at higher bias in all the data up to 85 K,
indicating that there is little contribution from the YBCO channel over this 
temperature range.
Similarly, we can argue that the thermal effects are minimized. The change of
curvature in Fig.~\ref{fig1}b at higher voltages around 60 K can be 
explained by the presence 
of vortices when the channel current approaches the critical current.
The most significant difference between the two sets of data 
is the structure near the zero bias in Fig.~\ref{fig1}b.
At higher temperatures it evolves into ZBCP,
but is completely absent for the smoother surface in 
Fig.~\ref{fig1}c. We attribute the 
possibility to observe ZBCP in our $c$-axis oriented
film to the known interfacial roughness of optimally grown YBCO
which facilitates $a-b$ plane transport. Consequently, such surface
morphology, due to the sign change of $d$-wave order parameter
would introduce non-vanishing weight of the $a-b$ plane Andreev
bound states~\cite{Fog97,Wal99}. To investigate this point we perform
control measurements with non-magnetic systems in the geometry\cite{size}
shown in Fig.~\ref{fig1}a where LSMO is replaced by LNO (Fig.~\ref{fig2}a)
and by Ag (Fig.~\ref{fig2}b). In spite of the significantly different electronic
properties of LNO and Ag, conductance data in both cases display similar
behavior: a ZBCP, present already at lowest examined temperature 
diminishes at higher temperature, consistent with the effects of thermal 
smearing.
These measurements (in particular the one on
LNO which, as a metallic oxide, is similar to LSMO) allow us
to investigate the
distinguishing conductance features arising from the spin-polarized transport.

In order to reveal the properties of interface transport from
Fig.~\ref{fig1}b near zero bias with greater clarity, we have to 
distinguish between the contributions of the two electrodes of the 
junction, to the G-V curves. 
Since we are mainly interested in studying the effect of spin polarized 
tunneling into the HTSC, we would like to 
remove the effect of the complicated density of states (DOS)
of the CMR electrode from the raw conductance data. The DOS of CMR
have been studied by tunneling spectroscopy~\cite{rajeev,amlan,wei} 
leading to  the corresponding conductance contribution at low bias
given by $G(V)=G_0(1+(|eV|/\Delta)^n),$ where $G_0$ is the zero bias value, 
$\Delta$ is the correlation gap and $0.5 < n < 1$.
To determine background conductance (assuming that it is predominantly due
to the CMR DOS), which should be removed from the measured G-V curves,
we first fit G(V) to the data at lowest T and next, for 
each higher T, apply  thermal smearing to the fitted curve\cite{high}.
The resulting curves, after removing background conductance at three different 
temperatures, are shown in
Fig.~\ref{fig3}a~\cite{rep}. Each curve is normalized with respect to its value at
$\Delta_0$, corresponding to the maximum gap~\cite{Dag00}. 
We compare these results with the theoretical analysis for transport across
CMR/HTSC junction by adopting the notation and  methods
from Ref.~\cite{Zut00}, generalized to finite temperature. The strength of
interfacial scattering is modeled by parameter Z$_0$  and the spin polarization 
is represented by X, the ratio of the exchange and Fermi energies for CMR. The 
limit X=0 depicts the unpolarized case, while X=1 corresponds to the complete 
polarization
of a half-metallic ferromagnet. To include the effects of different 
electronic densities in the two materials,
we use L$_0$, the ratio of Fermi wavevectors in HTSC and CMR\cite{L0}.
Additionally, to capture the main aspects of surface roughness, we average
results over  different interface orientations. This is
in contrast to the usually studied extreme cases of (110) and (100) planes,
corresponding to the maximum and minimum spectral weight of
the Andreev bound states, respectively.
The calculated conductance is normalized with respect to its value at the
maximum gap. 

For each G-V curve from Fig.~\ref{fig3}a, we plot in 
Fig.~\ref{fig3}b 
two curves for corresponding temperature with slightly different values
of $X$, to illustrate the sensitivity of results to the spin polarization
of a ferromagnet. We show that the the essential features from 
Fig.~\ref{fig3}a are well reproduced. The overall amplitude is expected to
be much smaller from measured values since in the analysis we do not include
$c$-axis contribution which could be modeled by a parallel conductance 
channel\cite{Dag00}. Findings from Fig.~\ref{fig3}b (which we have
also verified in a wide parameter range) show that at a fixed temperature the 
increase in the exchange energy reduces the amplitude of ZBCP. This
can partly explain smaller magnitude of ZBCP observed in LSMO/YBCO
as compared to the non-magnetic junctions. Another influence, contributing
to this difference in magnitude, could arise from the previously discussed
DOS effect of the CMR electrode.
Using the theoretical framework from Ref.~\cite{Zhu99,Zut00} it is possible
to understand various effects on conductance data  from the temperature
dependent exchange interaction responsible for the spin subband splitting 
and the spin polarization in the ferromagnetic region. 
In the limit of low T and almost complete spin polarization it is predicted 
that there would be a strong conductance suppression at low bias voltage 
where the transport properties are governed by Andreev 
reflection. For spin-polarized carriers only a fraction of 
the incident electrons from a majority spin subband will have partner from a 
minority spin subband in order to be Andreev reflected, leading to a reduced
charge transport across the interface~\cite{Jon95,Zhu99,Zut00,Zut99}.
If the spin-polarization decreases
with temperature there would be a smaller difference between population
in the two spin subbands resulting in an enhanced Andreev reflection and
more pronounced ZBCP. This temperature dependence is qualitatively different
from the effects of thermal smearing which reduce and broaden ZBCP.
Our raw data, from the blow up in Fig.~\ref{fig1}b,
provides therefore a strong support for the decreasing spin polarization 
with temperature which can even be sufficiently fast to off-set the opposite
effects of thermal smearing. Observing ZBCP at temperatures close to T$_c$
further suggests that fabricating our junctions have provided high
interfacial quality  without degrading the superconducting properties.
From the calculated conductance we
could infer high but not complete, spin polarization at lowest measured 
temperature, which is consistent with several other findings using
different techniques\cite{Sou98,Wor00}.
A detailed study on the LSMO/YBCO interface by x-ray magnetic 
circular dichroism~\cite{Sta99}  has concluded that the polarization
of LSMO is further suppressed, compared to the free surface, by the presence 
of a capping YBCO layer. In the case of LSMO, this should  serve as a caution 
for attempts to get the precise quantitative agreement in 
the degree of spin polarization obtained by various measurement techniques
which  often also involve different material fabrication.
The temperature dependence of the spin polarization from our data
is consistent with the results of spin-resolved 
photoemission spectroscopy data\cite{Par98}, which show that the spin 
polarization within 50 \AA~ of the free surface of LSMO drops faster
than the bulk with increasing temperature. 

In conclusion, by controlling the surface morphology in
a $c$-axis grown YBCO we have demonstrated qualitatively different
temperature evolution of the measured conductance features in the
ferromagnetic and non-magnetic superconducting heterostructures.
These differences are attributed to the suppression of the
spectral weight of the Andreev bound states by the spin-polarized
transport across the interface.

This work is supported by the US ONR Grants $\#$N000140010028, N000149810218, 
the NSF MRSEC Grant $\#$DMR-96-32521 and by Darpa. We thank I. 
Takeuchi, P. Fournier, I. I. Mazin and B. Nadgorny for valuable discussions.

\begin{figure}
\caption{a) Testing structure for LSMO/YBCO junction. b) Raw conductance
data at various temperatures as a function of applied voltage
for two different surface morphologies with the corresponding AFM images 
representing 5 $\times$ 5 $\mu$m scan size and same height range.
In b) optimally grown YBCO (T$_c$ = 90 K) and in c) smoother, 
off-optimal YBCO (T$_c$ = 83 K).}
\label{fig1}
\end{figure}

\begin{figure}
\caption{Temperature evolution of G-V data for a) Ag/YBCO and
b) LNO/YBCO junction. Both panels display ZBCP for a $c$-axis grown YBCO.}
\label{fig2}
\end{figure}

\begin{figure}
\caption{a) Normalized  conductance after removing the contribution of the
CMR DOS. 
b) Calculated results, averaged over all interfacial orientations.
For comparison, at each temperature in panel a), two curves
with slightly different spin polarization are shown.
The notation is following Ref.~\protect\cite{Zut00}.
Insets show curves fitted at the lowest examined temperature to include the
effects of the CMR DOS. Fitting parameters, as explained in the text,
are G$_0$=2.68 S, $\Delta$=0.058 meV, n=0.913 (left inset) for a faceted
surface, and G$_0$=0.62 S, $\Delta$=0.017 meV, n=0.743 (right inset)
for smooth surface.}
\label{fig3}
\end{figure}


\begin{references}
\bibitem{Jsi} Although the junction size we use is large compared to the
mean free path of the electrons, it was both shown 
experimentally ~\cite{Cov97,Saw99}, and 
predicted theoretically~\cite{Wal99} that a ZBCP can still be observed in
 the diffusive limit.
\bibitem{Off}  It is known that off-stoichiometric phases that may be
segregated into grain boundaries or the surface under optimized, 
well-equilibrated conditions cause large surface roughness. Under
off-optimal conditions, segregation is suppressed by the process of 
incorporation of defects into the lattice, which also leads to a
smoother surface.
\bibitem{field} We also performed these transport measurements with magnetic fields 
parallel to the interface. No significant anisotropy was observed for 
different directions of the applied field.
\bibitem{size} The actual pad size is 15 $\times$ 15 $\mu$m,
smaller than for the LSMO/YBCO junction, to obtain sufficiently large
signal.
\bibitem{high} For T above 40 K this procedure could not be simply implemented
because the emergence and strengthening of wing-like features at higher bias.
\end{references}
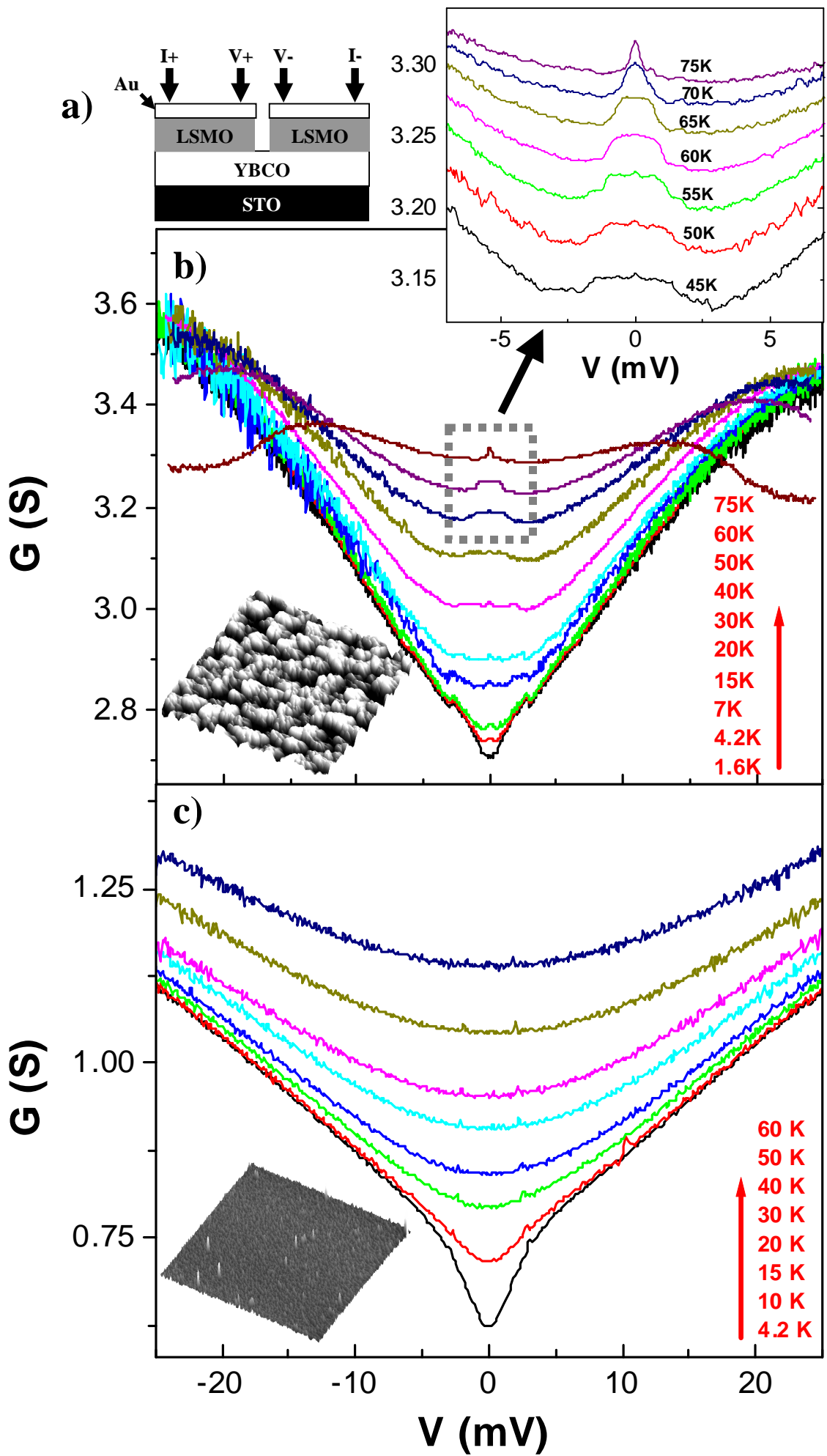
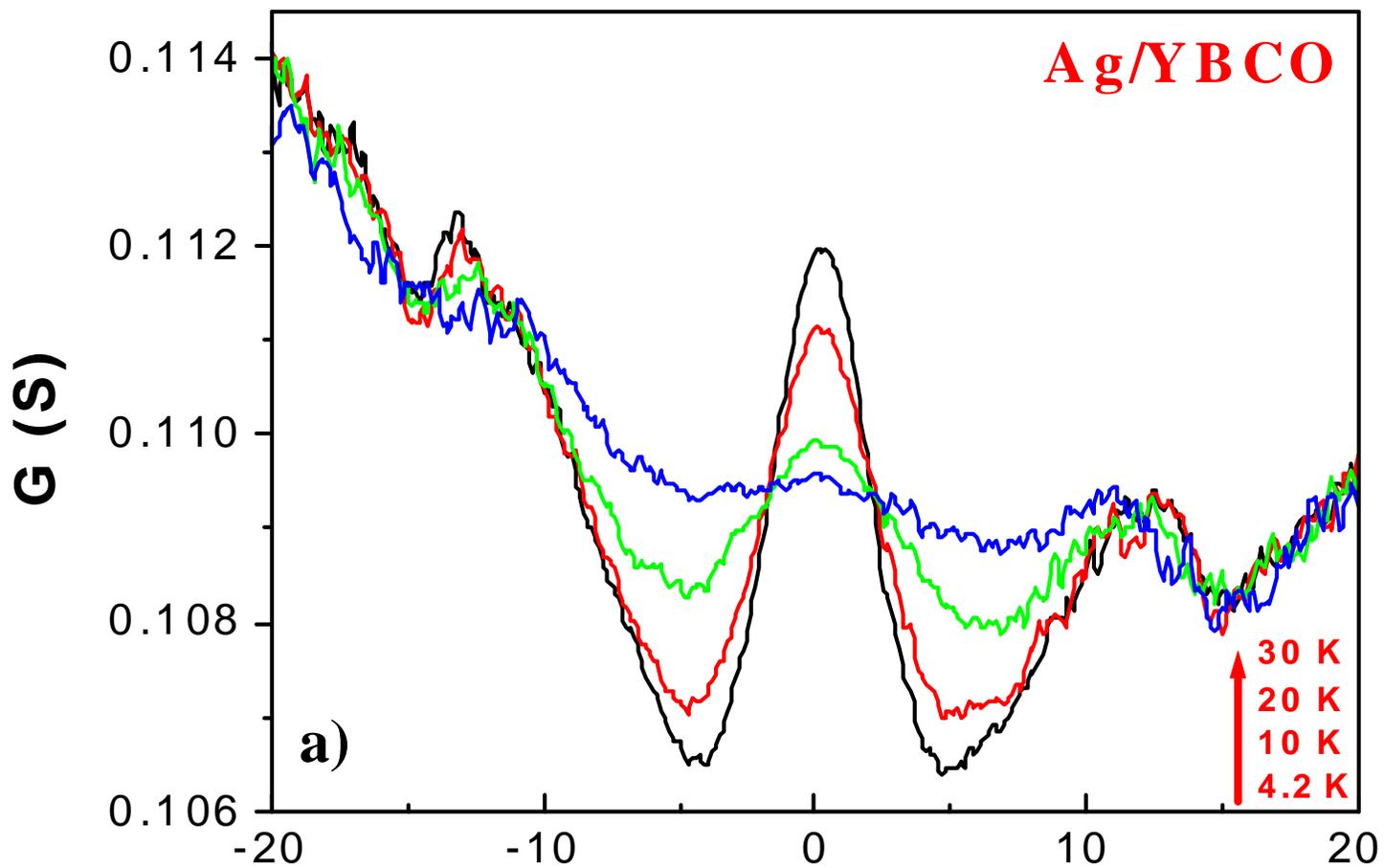
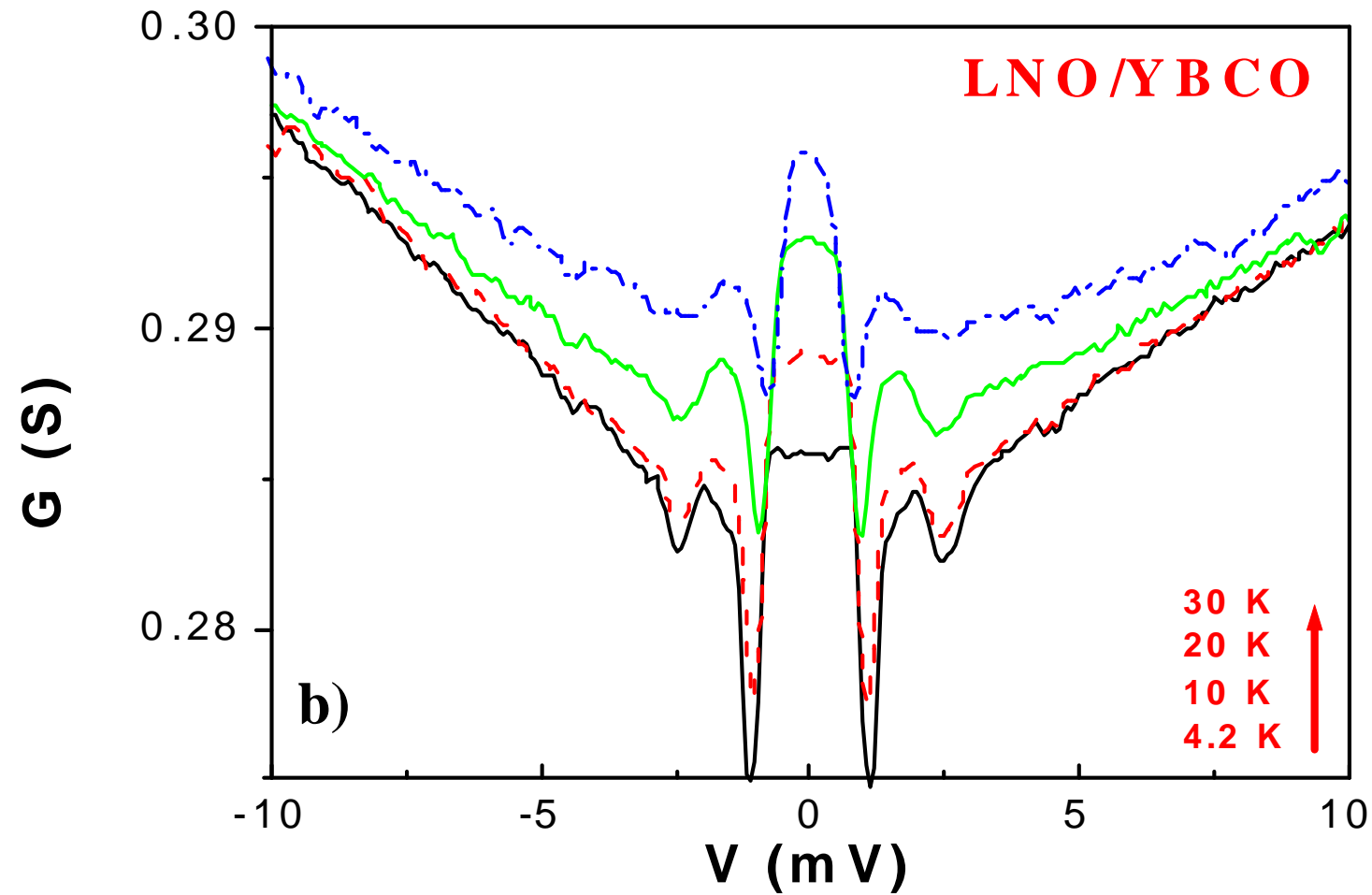
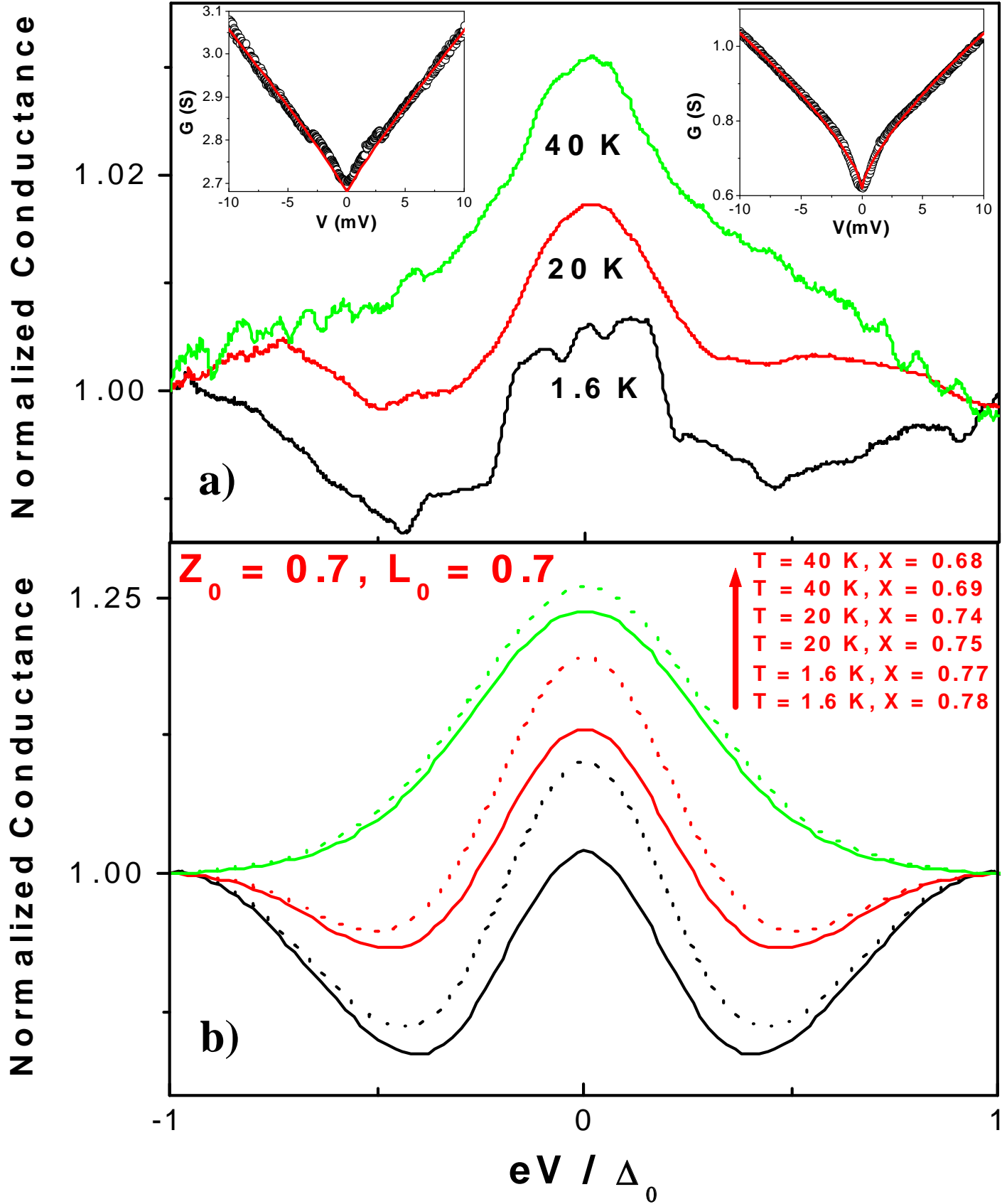
\end{document}